\documentclass[12pt]{article}

\pdfoutput=1

\usepackage{float}
\usepackage{hyperref}
\usepackage{authblk}
\usepackage{graphicx}
\usepackage[margin=0.5in]{geometry}

\def \TF    {T_{5/3}}
\def \TFB   {\bar{T}_{5/3}}
\def \pt    {p_{T}}
\def \Et    {E_{T}}
\def \Ht    {H_{T}}
\def \St    {S_{T}}
\def \met   {\not\!\!\Et}
\def \ttbar {\rm t\bar{t}}

\def \invfb {\rm {fb}^{\rm -1}}

\begin{document}

\title{\vspace{-15mm}\fontsize{24pt}{10pt}\selectfont\textbf{Search for top partners with charge 5e/3}}

\author[1]{A. Avetisyan}
\author[1]{T. Bose}

\affil[1]{Boston University}

\maketitle

\abstract{
A feasibility study of searches for top partners with charge 5e/3 at the upgraded Large Hadron Collider is performed. The discovery potential and exclusion limits are presented using integrated luminosities of 300~fb$^{-1}$ and 3000~fb$^{-1}$ at center-of-mass energies of 14 and 33 TeV.
}

\section{Introduction}
The recent discovery of a 125~GeV Higgs-like particle~\cite{higgspaper, Aad:2012tfa} has been a resounding success for the Large Hadron Collider (LHC) and its experiments. The focus now shifts to fully understanding the nature of electroweak symmetry breaking by measuring the properties of the Higgs boson and discovering new physics to address the critical questions still facing particle physics today. What is the solution to the hierarchy problem? What is responsible for dark matter? Why is gravity so weak?

A number of extensions to the Standard Model (SM) attempt to solve the above problems by predicting new particles. In particular, heavy partners of the top quark (``top partners'') arise in many of these models for addressing the hierarchy problem~\cite{DeSimone:2012fs, PhysRevD.81.075006, Dissertori:2010ug, Contino:2008hi}. Because of the large Higgs-top Yukawa coupling, radiative corrections to the weak scale from the top loop are considered the most significant source of fine-tuning in the SM. Natural theories of the weak scale must, therefore, include top partners whose interactions with the Higgs cancel the top-loop contribution to the weak scale.

These top partners are predicted to have masses close to the  electroweak symmetry breaking scale thus making them accessible at the LHC. In some cases they can also have exotic charge and contribute minimally to the coupling of the Higgs boson to gluons~\cite{Azatov:2011qy}. Searches for such top partners, therefore, continue to be viable despite the recent observation  of a 125 GeV SM Higgs-like resonance.

\begin{figure}[tp]
\begin{center}
\includegraphics[width=0.49\textwidth]{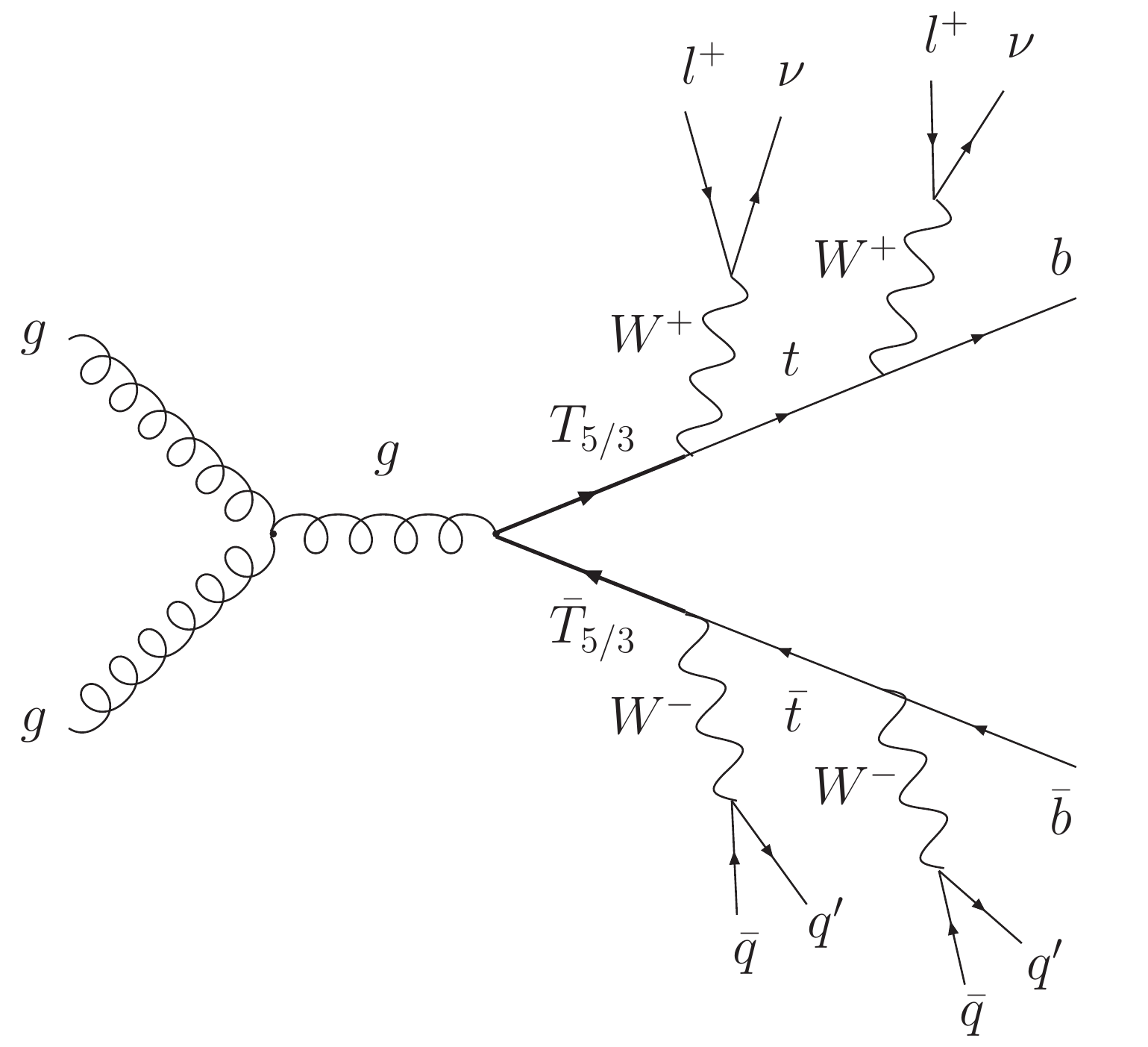} \hfill  
\includegraphics[width=0.49\textwidth]{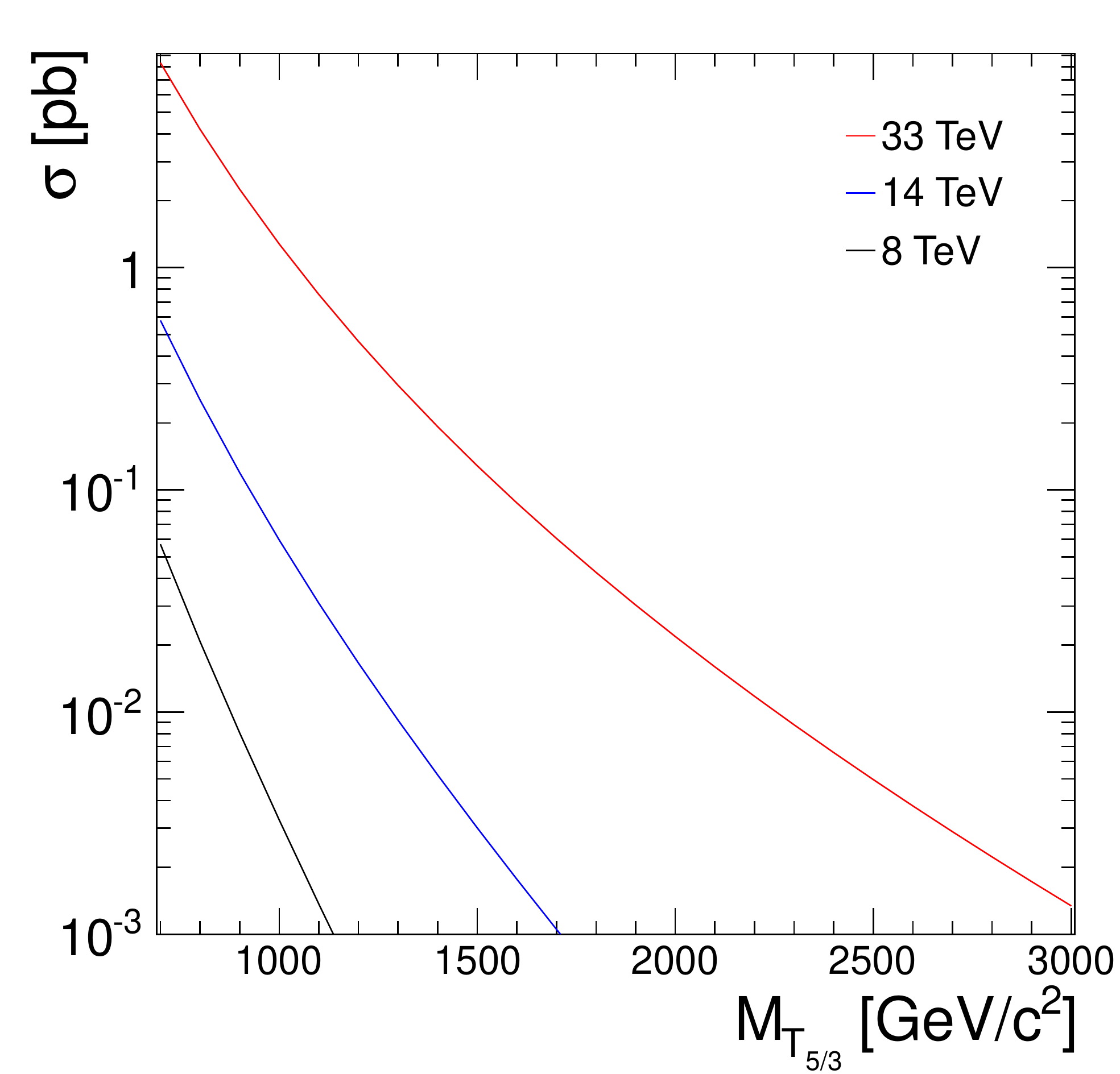} \\
\caption{The pair production and decay into same-sign dileptons of the $\TF$ (left) and the cross-section of the $\TF$ for
         various center-of-mass energies (right). The Feynman diagram is from~\cite{Contino:2008hi}. The cross-section is
         calculated using HATHOR~\cite{Aliev:2010zk}.}\label{fig:T53ProdAndDecay}
\end{center}
\end{figure}

We present a feasibility study for the $\mathrm{T_{5/3}}$, an exotic top partner with charge 5e/3 (where e is the charge of the positron) using simulated data with the Snowmass Combined LHC detector~\cite{Avetisyan:2013dta} at $\sqrt{s}=14$ and 33~TeV. We assume that the $\mathrm{T_{5/3}}$ is pair-produced and decays [see Fig.~\ref{fig:T53ProdAndDecay} (left)] via $\mathrm{T_{5/3}} \rightarrow tW^+$ and $t \rightarrow W^+ b$ (charge conjugate modes are implied throughout). We concentrate on the dilepton final state wherein the presence of same-sign leptons helps distinguish this process from the large $t\bar{t}$ background. Only contributions due to instrumental effects remain along with same-sign backgrounds with much smaller cross sections. 

\section{Signal and Background Samples}

All samples used in this study are generated using MadGraph~5~\cite{Alwall:2011st} and simulated using Delphes~3.0.9~\cite{Ovyn:2009tx,deFavereau:2013fsa}. For the backgrounds, the MadGraph generation is done in bins of 
the sum of transverse energy of all MadGraph-level particles in the event as described in~\cite{Avetisyan:2013onh}. The detector used as input to the simulation of both signal and background samples is the
Snowmass Combined LHC detector~\cite{Avetisyan:2013dta}. Two pileup scenarios are considered for each center-of-mass energy: one with 50 mean interactions per bunch crossing and one with 140 such interactions.

Signal samples are generated with the $\TF$ mass varied in intervals of 100~GeV. The mass in 14 TeV samples ranges from 0.7~TeV to 2.2~TeV while for 33 TeV samples, it varies from from 0.7~TeV to 3.0~TeV.
The principal same-sign backgrounds are $\ttbar W$, $\ttbar Z$ and the various combinations of diboson and triboson backgrounds (e.g. $W^{\pm}W^{\pm}$, $WZ$, $WWW$, $WWZ$, etc.).

\section{Event Selection}

In addition to having same-sign leptons, the decay of a $\TF\TFB$ pair also typically results in many jets and missing transverse energy ($\met$) from the neutrinos. At masses of the $\TF$ relevant at 
14 and 33 TeV, all of the $\TF$ decay products have transverse momenta ($\pt$) that are significantly higher than similar objects from any of the backgrounds. Therefore, the event selection is based
primarily on the $\pt$ of the decay products and sums thereof. The requirements at 14 and 33 TeV are listed in Table~\ref{tab:EvSel}. $\Ht$ is the scalar sum of all leptons and jets in the event with 
$\pt > 30$~GeV. $\St$ is the scalar sum of the $\Ht$ and the $\met$. 

\begin{table}[!Hhtb]
\centering
\caption{Event selection requirements as a function of center-of-mass collider energy.}
\begin{tabular}{|c|c|c|}
\hline
 Parameter              & 14~TeV Min [GeV] & 33~TeV Min [GeV] \\
\hline
Leading lepton $\pt$    & 80               & 150 \\ 
Second  lepton $\pt$    & 30               & 50  \\
Leading jet $\pt$       & 150              & 150 \\
Second  jet $\pt$       & 50               & 50  \\
$\met$                  & 100              & 200 \\
$\Ht$                   & 1500             & 2200 \\
$\St$                   & 2000             & 3000 \\  
\hline
\end{tabular}
\label{tab:EvSel}
\end{table}

Aside from the $\pt$ requirements, the number of decay products in the event is also a good discriminant between signal and background. However, it can be obscured by the fact that jets from 
highly boosted $W$ bosons and top quarks tend to merge into a single jet. To recover this information, we use the top and $W$ tagging algorithms implemented in the Delphes output~\cite{Avetisyan:2013dta}.
These algorithms use Cambridge-Aachen jets with a radius of 0.8 (CA8 jets). A CA8 jet is considered to be $W$-tagged if its mass is between 60 and 120 GeV and if the ``mass drop'' of the jet (the 
ratio of the leading sub-jet mass to that of the whole jet) is less than 0.4. Jets are top-tagged if their mass is between 140 and 230 GeV and if they have at least 3 sub-jets.

Given this implementation of jet substructure, the number of decay products is approximated by the number of ``constituents'' in the event. Each top-tagged CA8 jet counts as 3 constituents and 
each W-tagged jet counts as 2. All other jets in the event are reconstructed with the Anti-$k_{T}$ algorithm with a radius of 0.5 and are hence called AK5 jets. These jets are required to be
at least $\Delta R = \sqrt{(\Delta \phi)^{2} + (\Delta \eta)^{2}} > 0.8$ away from the $W$ and top-tagged jets where $\Delta \phi$ is the difference in azimuthal angle between the jets and 
$\Delta \eta$ is the difference in pseudorapidity ($\eta = -\ln{\left[\tan{\left(\frac{\theta}{2}\right)}\right]}$). Each AK5 jet counts as one constituent and so does each lepton with $\pt > 30$~GeV
except for the two leptons used for the same-sign requirement. A minimum of 5 constituents is required at both 14 and 33 TeV.

The signal and background yields of this selection are shown in Table~\ref{tab:BkgdYield} and~\ref{tab:SigYield} respectively.

\begin{table}[!Hhtb]
\centering
\caption{Background yields for 300~$\invfb$.}
\begin{tabular}{|c|cc|cc|}
\hline
                 & \multicolumn{2}{c|}{14 TeV}  &  \multicolumn{2}{c|}{33 TeV} \\
\hline
	         &    50 Pileup &    Pileup 140	&  Pileup 50	&  Pileup 140  \\
\hline
Tribosons	 &	0.53	&	0.96	&	1.45	&	3.56\\
Diboosns	 &	3.66	&	6.94	&	6.24	&	17.37\\
$t\bar{t}$+Boson &	5.70	&	6.80	&	8.57	&	12.43\\
\hline
Sum      	 &	9.88	&	14.70	&	16.26	&	33.36\\
\hline
\end{tabular}
\label{tab:BkgdYield}
\end{table}

\begin{table}[!Hhtb]
\centering
\caption{Signal yields for 300~$\invfb$.}
\begin{tabular}{|c|cc|cc|}
\hline
                 & \multicolumn{2}{c|}{14 TeV}  &  \multicolumn{2}{c|}{33 TeV} \\
\hline
$\TF$~Mass (TeV) &    50 Pileup &    Pileup 140 &  Pileup 50	&  Pileup 140  \\
\hline
0.7	&	270.47	&	293.83	&	870.72	&	951.17\\
0.8	&	222.80	&	240.15	&	648.39	&	791.04\\
0.9	&	164.38	&	169.71	&	533.27	&	708.37\\
1.0	&	120.48	&	131.80	&	466.74	&	536.83\\
1.1	&	83.65	&	95.69	&	389.30	&	527.08\\
1.2	&	56.77	&	64.44	&	414.78	&	518.18\\
1.3	&	36.84	&	43.05	&	282.59	&	386.26\\
1.4	&	21.31	&	25.14	&	253.73	&	319.38\\
1.5	&	13.29	&	15.89	&	194.14	&	255.87\\
1.6	&	8.12	&	9.59	&	139.97	&	197.43\\
1.7	&	4.71	&	5.72	&	101.48	&	151.23\\
1.8	&	2.81	&	3.48	&	73.96	&	108.77\\
1.9	&	1.76	&	2.12	&	60.33	&	82.41\\
2.0	&	0.99	&	1.24	&	42.03	&	63.76\\
2.1	&	0.63	&	0.74	&	33.49	&	49.09\\
2.2	&	0.36	&	0.43	&	22.56	&	37.72\\
2.3	&	-	&	-	&	19.99	&	27.87\\
2.4	&	-	&	-	&	13.97	&	21.60\\
2.5	&	-	&	-	&	10.20	&	14.82\\
2.6	&	-	&	-	&	8.36	&	13.08\\
2.7	&	-	&	-	&	6.07	&	8.71\\
2.8	&	-	&	-	&	4.53	&	7.01\\
2.9	&	-	&	-	&	3.23	&	5.35\\
3.0	&	-	&	-	&	2.76	&	4.28\\
\hline
\end{tabular}
\label{tab:SigYield}
\end{table}

\section{Discovery Potential}

\begin{table}[t]
\centering
\begin{tabular}{|c|c|c|c|c|c|}
\hline
Collider      &	Luminosity   & Pileup & 3$\sigma$ evidence & 5$\sigma$ discovery & 95\% CL  \\ \hline

LHC 14 TeV    &	300 fb$^{-1}$ &    50  & 1.51 TeV           & 1.39 TeV            & 1.57 TeV \\ \hline
LHC 14 TeV    &	300 fb$^{-1}$ &   140  & 1.50 TeV           & 1.38 TeV            & 1.58 TeV \\ \hline
LHC 14 TeV    &	3 ab$^{-1}$   &    50  & 1.67 TeV           & 1.57 TeV            & 1.76 TeV \\ \hline
LHC 14 TeV    &	3 ab$^{-1}$   &   140  & 1.66 TeV           & 1.55 TeV            & 1.76 TeV \\ \hline \hline
LHC 33 TeV    & 300 fb$^{-1}$ &    50  & 2.36 TeV           & 2.13 TeV            & 2.48 TeV \\ \hline
LHC 33 TeV    & 300 fb$^{-1}$ &   140  & 2.17 TeV           & 2.15 TeV            & 2.47 TeV \\ \hline
LHC 33 TeV    & 3 ab$^{-1}$   &    50  & 2.61 TeV           & 2.40 TeV            & 2.77 TeV \\ \hline
LHC 33 TeV    & 3 ab$^{-1}$   &   140  & 2.50 TeV           & 2.35 TeV            & 2.69 TeV \\ \hline

\end{tabular} \hspace{-0.138cm} 
\vspace{0.3cm}
\caption{Expected mass sensitivity for charge $5/3$ pair production with decay into $t W$.}
\label{tab:Results}
\end{table}%

Based on the event yields of the above selection, the significance as a function of $\TF$ mass is computed at the various integrated luminosity, pileup and center-of-mass energy scenarios. In the 
absence of a signal, limits on the $\TF$ mass are also computed. In agreement with other Snowmass top-related searches, the systematic uncertainty on all of the backgrounds is assumed to be 20\%. 
The significances and limits are shown in Table~\ref{tab:Results}.

\section{Mass Reconstruction}

In the event that the $\TF$ is discovered, it can be distinguished from models with similar signatures by using the mass distribution of the $\TF$. The mass can be fully reconstructed when the decay of
one quark in the $\TF\TFB$ pair is fully leptonic while the decay of the other is fully hadronic. The hadronic decays result in 2 partons from the W boson and 3 partons from the top quark. The 
reconstruction proceeds by reconstructing the Lorentz vectors of the W boson and top quark and then combining them to construct the $\TF$. If the event has W-tagged or top-tagged CA jets, these are assumed to
be the corresponding particle. If there are not enough tagged CA jets to reconstruct the $\TF$, the missing particles are reconstructed using AK5 jets. A W boson reconstructed from AK5 jets must be
within 20~GeV of the W mass whereas a top quark reconstructing using AK5 jets must be within 30~GeV of the top quark mass. If there are more top quark or W boson candidates in an event than necessary 
to reconstruct the $\TF$, the ones closest to the expected mass are used and the rest are discarded. 

\begin{figure}[!Hhtb]
\begin{center}
\includegraphics[width=0.49\textwidth]{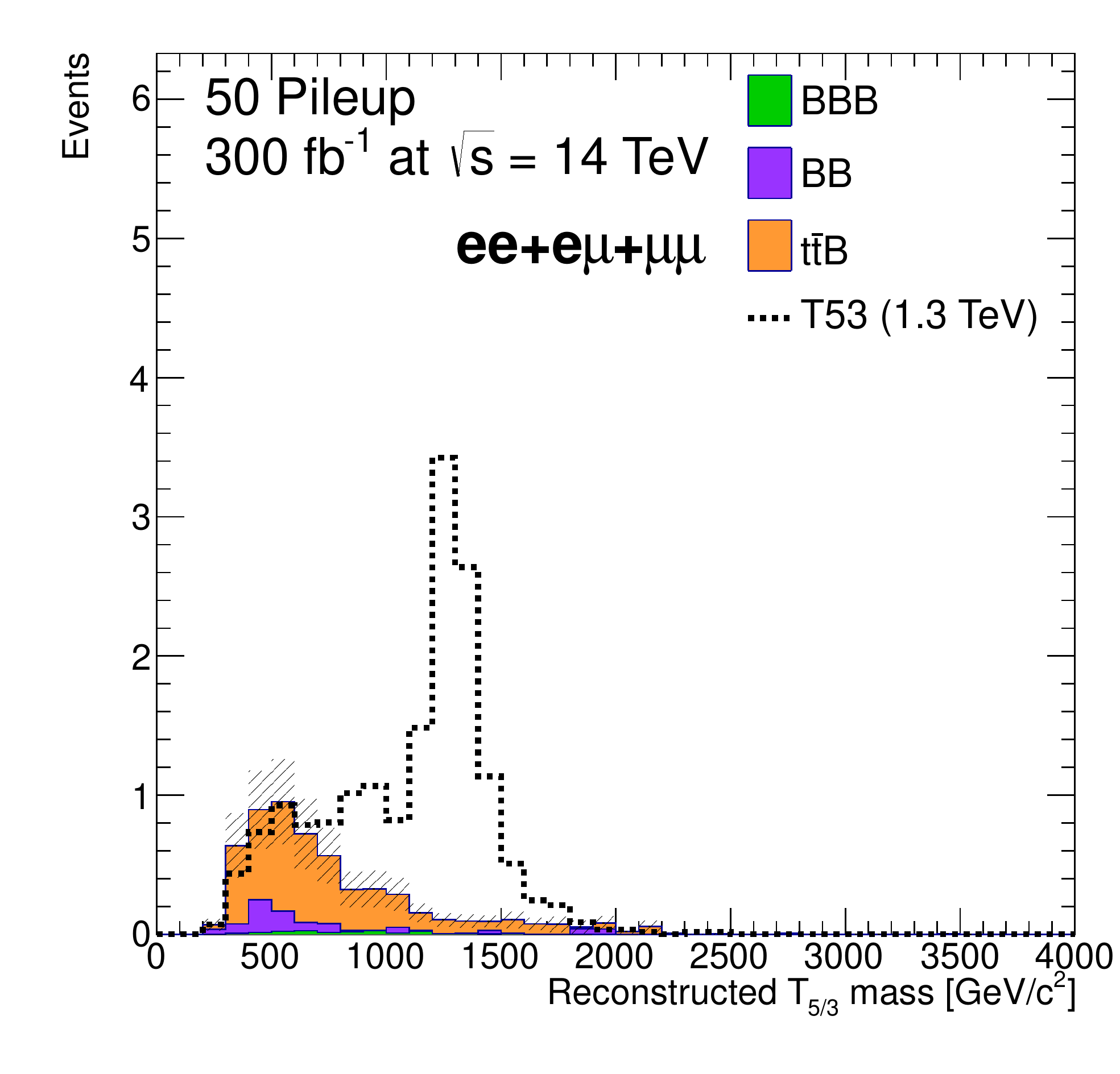} \hfill  
\includegraphics[width=0.49\textwidth]{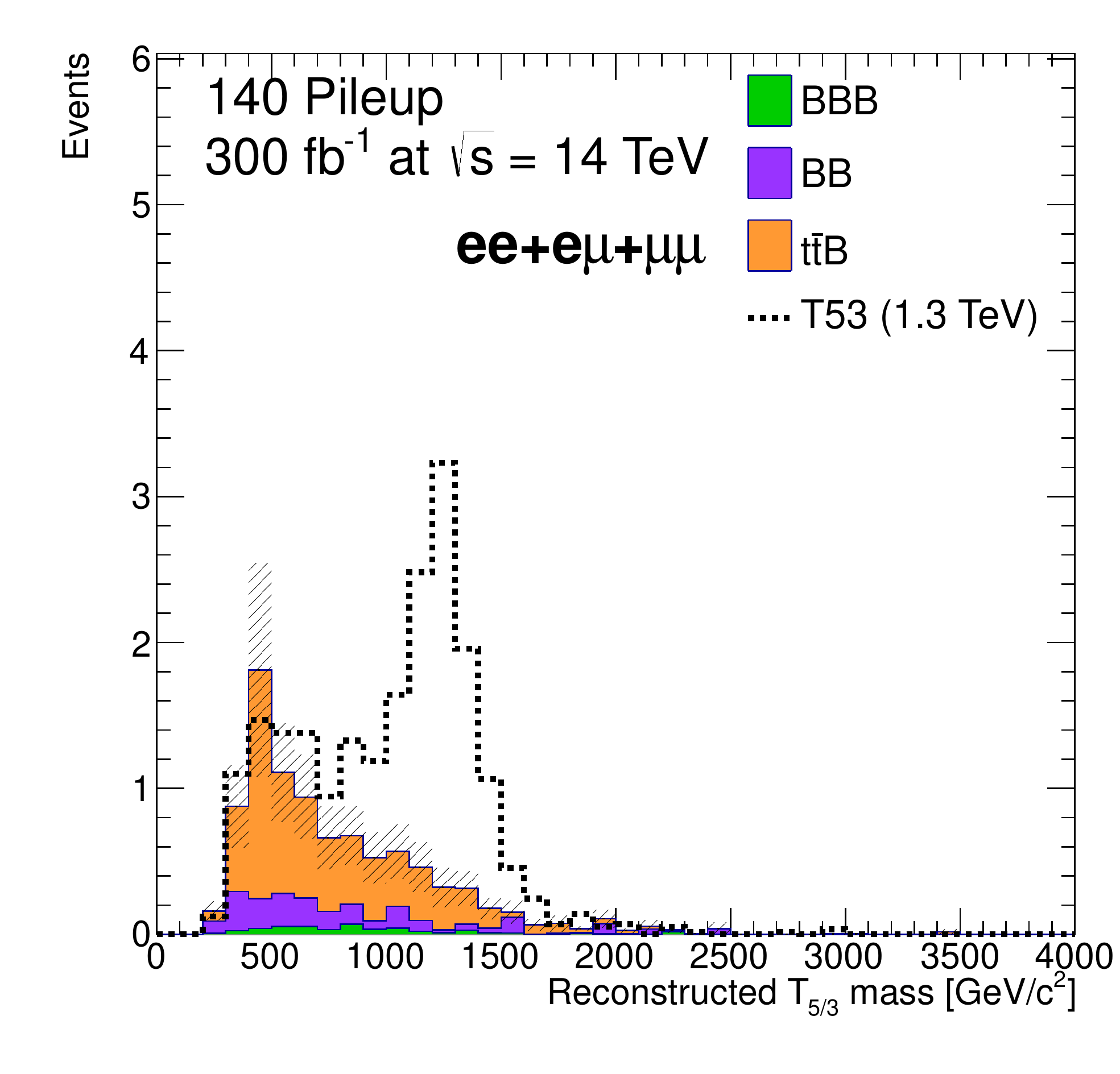} \\
\caption{The reconstructed mass distribution at 14~TeV with 50 (left) and 140 (right) pileup.}\label{fig:MassReco}
\end{center}
\end{figure}

The selection used prior to the mass reconstruction is the same as the full event selection, but without the $\met$ and $\St$ requirements. Instead of the latter, the $\TF$ mass is required to be greater
than the transverse mass of the two leptonic W bosons. The possible jet combinations in the preferred order of reconstruction are shown in Table~\ref{tab:MassRecoPref}. The distributions of the 
reconstructed mass at 14 TeV with 50 and 140 pileup are shown in Figure~\ref{fig:MassReco}. For $\TF$ masses accessible at 33~TeV, the W bosons and top quarks are boosted beyond the $\pt$ range of
current W and top tagging algorithms. Reconstruction of the $\TF$ at 33~TeV would make use of improved detectors and improved tagging algorithms as described in~\cite{Calkins:2013ega}.

\begin{table}[!Hhtb]
\centering
\caption{Possible jet combinations used to reconstruct the $\TF$ in the order of preference used for the reconstruction.}
\begin{tabular}{|c|c|c|}
\hline
 top-tagged jets &    W-tagged jets &  AK5 jets \\
\hline
 1               &     1            &  0        \\ 
 1               &     0            &  2        \\   
 0               &     2            &  1        \\  
 0               &     1            &  3        \\  
 0               &     0            &  5        \\  
\hline
\end{tabular}
\label{tab:MassRecoPref}
\end{table}

\section{Conclusion}

We have performed a feasibility study of searches for top partners with charge 5e/3 at the upgraded Large Hadron Collider. At 14 TeV, it is possible to discover such top parters with masses
up to 1.6~TeV or exclude masses below 1.8~TeV. At 33 TeV, the discovery potential increases to 2.4 TeV and the exclusion grows to 2.8 TeV. 

\bibliographystyle{utphys}
\bibliography{T53SnowmassWhitepaper}

\end{document}